\documentclass[11pt]{article}
 \usepackage{amssymb}
 \usepackage[round]{natbib}
 \usepackage{bm}
 \usepackage{color}
 \usepackage{graphicx}
 \newtheorem{theorem}{Theorem}
 \newtheorem{lemma}[theorem]{Lemma}

 \usepackage{xspace}
 \usepackage{setspace}
 \newcommand{\pin}{\ensuremath p_{{\rm inv}}\xspace}
 
 \title{Twisted trees and inconsistency of tree estimation when gaps are treated as missing data
 -- the impact of model mis-specification in distance corrections} 
 \author{Emily Jane McTavish$^{1,2,\ast}$
         \and
         Mike Steel$^{3}$
         \and
         Mark T.~Holder$^{1,2}$}
 
\begin{document}
 
 \maketitle
 
 \noindent1. Department of Ecology and Evolutionary Biology, University of Kansas, Lawrence KS, USA\\
 \noindent2. Heidelberg Institute for Theoretical Studies, Heidelberg, Germany\\
 \noindent3. Biomathematics Research Centre, University of Canterbury, Christchurch, New Zealand\\
 \noindent$\ast$ to whom correspondence should be addressed. Email: ejmctavish@gmail.com.
 \begin{abstract}

Statistically consistent estimation of phylogenetic trees or gene trees
    is possible if pairwise sequence dissimilarities can be converted to a set of
    distances that are proportional to the true evolutionary distances.
\citet{susko_inconsistency_2004} reported some strikingly broad results about
    the forms of inconsistency in tree estimation that can arise if
    corrected distances are not proportional to the true distances.
They showed that if the corrected distance is a concave function
    of the true distance, then inconsistency due to long branch attraction
    will occur.
If these functions are convex, then two ``long branch repulsion'' trees
    will be preferred over the true tree -- though these two incorrect trees
    are expected to be tied as the preferred true.
Here we extend their results, and  demonstrate the existence of a tree shape (which 
    we refer to as a ``twisted Farris-zone'' tree) for
    which a single incorrect tree topology will be guaranteed to be preferred
    if the corrected distance function is convex.
We also report that the standard practice of treating gaps in sequence alignments
    as missing data is sufficient to produce non-linear corrected distance functions
    if the substitution process is not independent of the insertion/deletion process.
Taken together, these results imply inconsistent tree inference under mild conditions.
For example, if some positions in a sequence are constrained to be free of substitutions and 
    insertion/deletion events while the remaining 
    sites evolve with independent substitutions and insertion/deletion events, 
    then the distances obtained by treating gaps as missing data can support an
    incorrect tree topology even given an unlimited amount of data.
\end{abstract}

\section{Keywords}
phylogenetics, distance methods, inconsistency, invariant sites, insertion, deletion, gaps as missing data

\section{Introduction}
Distance-based methods are fast and statistically consistent estimators of tree topology 
if the input distances converge (with increasing data) to values that are  proportional to the evolutionary distance between tips.
An evolutionary distance is the number of substitution events that have occurred along the path separating two tips.
Typically a distance correction procedure is applied to the observed sequence
    differences to obtain a more accurate estimate of the evolutionary distance
    between pairs of sequences.
However, in many cases it is not possible to correctly account for the evolutionary processes
    which generated the data. 
In other words, it is not always possible to accurately estimate
    the evolutionary distance for pairwise measurements of dissimilarity.

In a pioneering paper, \citet{susko_inconsistency_2004} showed how model misspecification can lead to transformed 
evolutionary distances that are non-linear with respect to evolutionary distance
    (i.e. concave or convex), and for which there are regions of tree space for which neighbor joining will be inconsistent.  
We extend this result further (Theorem~\ref{mainThm} in Appendix A)  by showing how virtually all misspecified correction functions lead to  (strong) inconsistency (an incorrect tree will be unambiguously favoured by neighbor-joining).  A main focus of this paper involves a particular study of model misspecification in  distance corrections that treats gaps as missing data.

\section{Model}
For variants of the simplest model of sequence evolution \citep{jukes_evolution_1969}, all nucleotides are equally exchangeable and
    the simple proportion of sites that differ, the $p$-distance, is a sufficient statistic for estimating 
    an evolutionary distance.
    Under such a model, $M_g$, the expected $p$-distance between a pair of taxa is a function of the evolutionary distance  (path length in the tree) $t$ between the taxa, that is, we have $\mathbb{E}_g[p]=g(t)$, where the function $g$ is a monotonically (strictly) increasing function of $t$  which is analytic (i.e. has a power series expansion, and so derivatives exist of all orders) and satisfies $g(0) = 0$. For example, for the Jukes-Cantor model we have $g(t) =\frac{3}{4}\left(1 - e^{-\frac{4}{3}t}\right)$.
If the distances are corrected under a (possibly different), fully exchangeable model, $M_f$, then the estimated
    evolutionary distance $\hat{t}$ is usually computed from the $p$-distance by simply using the `plug-in' formula $\hat{t} = f^{-1} (p)$.

Thus, for any generating model for which $p$ converges in probability towards its expected value $\mathbb{E}_g[p] = g(t)$ (e.g. i.i.d. site substitution models) the estimated evolutionary distance $\hat{t}$ will converge towards
$\overline{t}=h(t)$, where $h(t) = f^{-1}(g(t))$.
Note here that both $p$ and $\hat{t}$ are random variables, while $\overline{t}$ is simply a function of $t$.
Notice that this `transformed' evolutionary distance $\overline{t}$ is not exactly the expected value of $\hat{t}$,  even when $f=g$ \citep{tajima93}, since the expectation of a non-linear function of random variable is not generally equal to the function evaluated at the expected value of that variable.  Nevertheless, for any i.i.d. site substitution model, 
the difference between $\overline{t}$ and the expected value of $\hat{t}$ decays towards zero as the sequence length grows.

Notice also that  when  $f=g$ (i.e. the correction model matches the generating model) then 
$\overline{t}=t$.  However, in general, $\overline{t}$ need not be equal to $t$, except when $t=0$.
For example, if the generating model is the Jukes-Cantor model with some form of among-site
    rate heterogeneity and the correcting model that does not assume the same form of rate heterogeneity
    then $\overline{t}$ can depend on  $t$ in a quite non-linear way \citep{soubrier2012}.  
    
    In this paper we are interested in determining when the transformed evolutionary distances $\overline{t}$ will favour a different tree to the tree generating the data. 
    In particular, we explore an example of how ignoring the process of insertion and deletion (referred to 
    jointly as indels hereafter) can lead
    to statistical inconsistency in an otherwise correctly modeled substitution process.
Inconsistency occurs in this case even
    when the alignment of residues is correct.

\begin{figure}[ht]
\begin{center}
\includegraphics[scale=2.0]{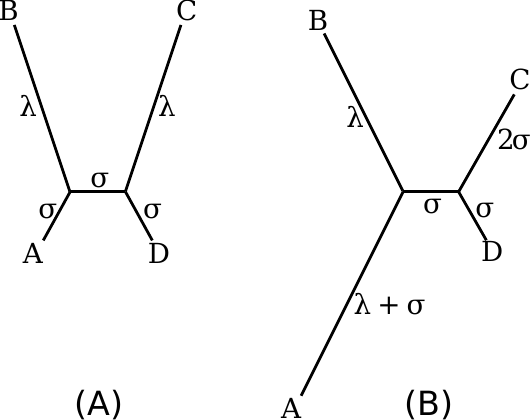}
\caption{(A) The Felsenstein-zone tree with branch lengths used in the proof of Lemma~\ref{concaveThm}; (B) The `twisted' Farris-zone tree used in the proof of Lemma~\ref{convexThm}.}
\label{figcom}
\end{center}
\end{figure}

\cite{susko_inconsistency_2004} studied general properties of $\overline{t}$ as a function of $t$. 
If this function is linear (i.e. when the correction model matches the generating model up to a constant factor) then distance-based tree estimation will be statistically consistent.
If the function is concave, inference can be inconsistent and positively misleading due to long branch attraction.
They also show that if the function is convex, two long branch repulsion trees are expected to be equally preferred over the correct tree.
In Appendix A we establish a more general result:  outside of the special case where the correcting generating model matches the generating model  up to a constant factor, there will always exist tree shapes
    for which neighbor-joining will estimate a single incorrect tree from $\overline{t}$. 
The tree shapes used to demonstrate this result are the familiar Felsenstein-zone tree \citep[Fig. \ref{figcom}A;][]{felsenstein_cases_1978}
and a tree that we refer to as the ``twisted Farris-zone'' tree (Fig. \ref{figcom}B).

\subsection{The gaps as missing data convention}
It is common practice to treat a gap in a sequence as missing data in
    phylogenetic estimation based on distances, parsimony scores or likelihoods.
In the context of a pairwise distance calculation, this treatment means that
    positions with a gap in either sequence are disregarded because
    they cannot be counted as either a similarity or a difference.
Omitting indels from distance corrections obviously forfeits the opportunity
    for  learning about the evolutionary distance from insertions and deletion
    events.
However, one may hope that treating sites with gaps as missing data would not perturb
    a distance estimate that relies solely on substitution events.
If the substitution and indel processes are completely independent, this is the case.

Consider the case of sequences that are generated by:
    a time-reversible stochastic process of insertions and deletion,  and
    a model of substitutions for which there is a statistically consistent distance correction.
If the alignment is known without error, then the only effect of the indel process is to
    introduce a fraction of sites, $z$, for which one sequence lacks a residue and the other sequence has a residue.
These are the gapped positions in a pairwise alignment.
Note that the presence of gap in a column in the alignment is not handled by deleting the
    column.
The gap only affects pairwise comparisons involving a sequence which contains a gap.
A full description for $z$ for a full alignment would require some additional notation to indicate which sequences
    are being compared.
Our argument below applies to any pairwise distance, so we simply use $z(t)$ to
    describe the expected proportion of gapped position in any pairwise distance for
    sequences separated by path length, $t$.

The fraction of gapped positions will be a function of the evolutionary distance with: $z(0) = 0$ because
at no distance there are no opportunities for indels, and $z(t) < 1$ for all $t$.
The latter property can be seen by treating one of the two sequences as if it were the ancestral sequence.
This is permissible because we have assumed that the indel process is time reversible.
The probability of a residue surviving from the ancestral sequence to the descendant sequence is described
    by an exponential function with rate parameter controlled by the rate of deletions.
This probability remains $>0$ for all values of the evolutionary distance, hence there is a non-zero probability of an ungapped
    position, and $z(t)$ cannot equal 1.

In a typical consistency proof, we consider sequences of ever increasing length. 
We note that indel models (e.g. the TKF91 model) imply a equilibrium sequence length.
Here we discuss statistical consistency by considering what happens as the number of loci increases
    without bound, but the equilibrium length of each locus is determined by the parameters of the 
    indel model.
Hence the total sequence length approaches infinity, while it is still appropriate to describe 
    the sequence as being generated by the indel process.

For the standard substitution models, we can consistently estimate the distance
    if the indel process has insertion and deletion rates of 0.
In this case there are no gapped columns and $z(t) = 0$.
In the more general case, if we only consider site patterns in which no gaps occur, the probability of a site pattern $s$
    for branch length $t$ is $\Pr(s|t) = (1-z(t)) \Pr'(s|t)$ 
    where $\Pr'(s|t)$ is the usual site pattern probability (when we have no missing data caused by gaps), 
    and $(1-z(t))$ is the probability of not containing a gap.
The multiplication of the probability of not containing a gap by the probability of the site pattern 
    given the branch length is valid whenever the substitution and indel process are independent.
We can see that calculating the probability of each ungapped site pattern by using the fraction of
    ungapped sites that display the pattern will result in $\Pr'(s|t)$ because this
    will constitute dividing the probability of each pattern by $(1-z(t))$.
Thus the spectrum of ungapped pattern frequencies will converge to exactly the same frequencies of
    the patterns when there are no indel events.

Thus, if the indel process and substitution process are independent, treating gaps as missing data will 
    not cause statistical inconsistency of distance-based tree inference.
Note that this result does not contradict the proof by \citet{Warnow2012} that treating
    gaps as missing data can lead to inconsistency in maximum likelihood.
Her proof focuses on the maximum likelihood criterion and is restricted to the case in
    which internal branch lengths for the substitution process are equal to 0 (there are 
    no substitution events).
Internal branch lengths of 0 lead to inconsistency without the complication of an indel process.
Our result applies to cases in which the tree method is capable of consistently estimating
    the tree if there are no indels.

\subsection{Cases in which indel processes and substitution process are not independent}
If the occurrence of an indel affects the probability of a substitution, then the previous argument does not hold.
In fact, we cannot use the argument above under any violation of the independence assumption.
For example, if some subset of sites is constrained by evolution and thus free of both substitutions and 
indels, then it is possible for the gaps-as-missing-data convention to lead us to the wrong tree.
In such cases, the gapped sites are a biased sample with respect to the substitution process.
See \citet{roure_impact_2013} for a discussion of other contexts in which non-random patterns of missing data perturb phylogenetic estimation.
Specifically, if the distribution of missing sites is not independent of the evolutionary rates at those sites
   this bias can lead to problems in phylogenetic reconstruction \citep{grievink_missing_2013, roure_impact_2013}.

\subsection{Paired invariants model}
Consider the case of sequences being generated under the TKF91 \citep{thorne_evolutionary_1991} indel model and the Jukes-Cantor (JC) \citep{jukes_evolution_1969} substitution model, but
    with invariant sites.
In particular, let the ``paired invariant sites'' model refer to the case in which some fraction of sites are
    free from both indels and substitutions and the other parts of the sequence are described by the TKF91 and JC models.
In terms of the formalism of the TKF91 model, this would require that the each invariant site which is followed by a
    region that is free to vary is considered to have a new ``immortal link'' to the right of the invariant site.
We consider the case in which alignment is known without error.

Let $\pin$ denote the expected proportion sites in a sequence which are invariant with respect to indels and substitutions.
In the TKF model single nucleotide insertions and deletions can occur at any site in the alignment \citep{thorne_evolutionary_1991}.
Under the TKF model, at equilibrium the expected rate of insertions per locus is equal to the expected rate of deletions per locus.
The TKF model is usually described with insertion rates per link and deletion rates per link.
In that parameterization the insertion and deletion rates can differ.
We call the deletion rate scaled per nucleotide $\theta$.

When computing a pairwise distance, the gaps-as-missing-data correction removes sites in which either 
    sequence has a gap from consideration.
The expected length of a locus under the paired invariants model will be denoted $N$.
This will be a function of the expected length of each block of variable sites, which is
    a function of the insertion rate relative to the deletion rate.
Our argument applies to any insertion rate which leads to a non-infinite equilibrium sequence length.
So we phrase the argument in terms of the per-locus expected length and do not use the insertion
    rate parameter explicitly in our argument.

Under the TKF91 model, each block of variable sites is expected to follow a geometric distribution with a parameter that depends
    on the ratio of the per-link insertion and deletion rates.
Because sites with an insertion and then a deletion are typically culled from an alignment, we consider a
    pairwise alignment length to be the length of the correct alignment after all positions with gaps in both
    members of the pair are removed.
Even though the expected number of nucleotides in each sequence does not change, the insertion of new positions and 
deletion of sites means that the pairwise alignment length grows as a function of the evolutionary distance.
In the paired invariants model, let $a(t,\theta,\pin)$ denote the expected length of 
    a pairwise alignment of two sequences separated by path length, $t$.
Then:
\begin{eqnarray}
 a(0,\theta,\pin) &=& N\\
\lim_{t \rightarrow \infty}  a(t,\theta,\pin) &=& N(\pin  + 2(1-\pin))
\end{eqnarray}
where $N\pin$ is the number of invariable columns in the alignment.
$N(1-\pin)$ columns are expected to be in the ancestor but deleted along the path to the descendant.
Because the process started at equilibrium, we expect them to be replaced by $N(1-\pin)$ inserted sites.

For each site that is free to vary in the ancestor, 
    the probability that it survives to the descendant is $e^{-\theta t}$, using the exponential distribution.
We refer to columns where there is a nucleotide in both the ancestor and the descendant as ``ungapped columns''.
The expected number of ungapped columns is $$b(t,\theta,\pin) = N\left(\pin + (1-\pin)e^{-\theta t}\right)$$
Note that $\lim_{t \rightarrow \infty} b(t,\theta,\pin)=N \pin$.

\begin{figure}
\caption{Properties of the paired invariants model with $\pin=0.2$ and $\theta=0.1$.
A. The proportion of aligned sites which are free to vary as a function of time (Eqn. \ref{propUngappedVar})
B. Pairwise nucleotide substitution distance through time (Eqn. \ref{expPPaired}).
Note 5-fold difference in $t$-axis scale between A and B.}
\includegraphics[scale=0.7]{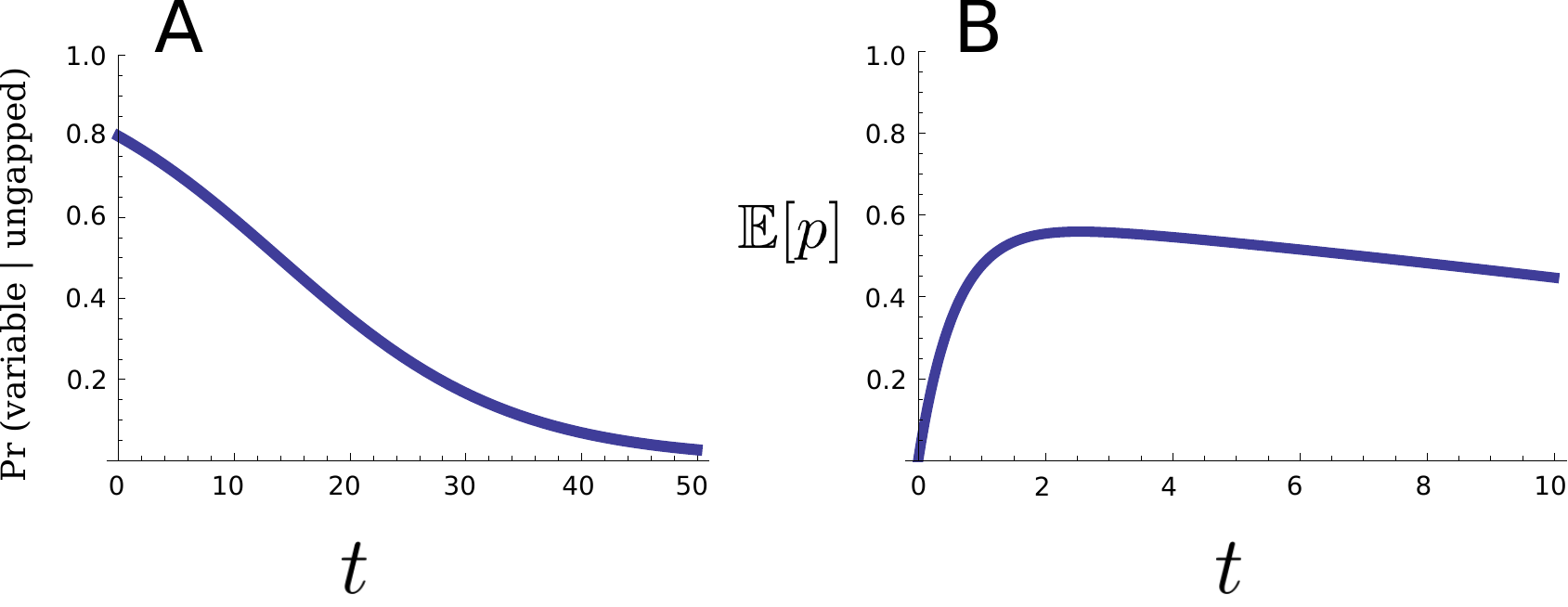}
\label{figgen}
\end{figure}

The expected proportion of {\em residues} in a sequence which are free to vary remains constant at $1-\pin$
    as branch length approaches infinity.
However, if we consider only ungapped columns in the true alignment of two sequences,
    we see that the proportion of these sites which are variable approaches 0 as deletions
    continue to reduce the number of aligned columns among the class of variable sites.
The expected proportion of ungapped columns that are free to vary is:
\begin{eqnarray}
    \Pr(\mbox{variable}\mid\mbox{ungapped}) & = & \frac{N (1-\pin)e^{-\theta t}}{b(t,\theta,\pin)} \nonumber\\
        & = & \frac{(1-\pin)e^{-\theta t}}{\pin + (1-\pin)e^{-\theta t}} \label{propUngappedVar}
\end{eqnarray}
This function is plotted in Fig.  \ref{figgen}{A} for the case when $\pin=0.2$ and $\theta=0.1$.

Recall that under the Jukes-Cantor model the probability of a site having a different nucleotide
from its ancestor across a path of length $t$ is $\frac{3}{4}\left(1 - e^{-\frac{4}{3}t}\right)$.
For the Jukes-Cantor model with invariant sites the probability of a difference, conditional on a site 
being a member of the variable class is:
\begin{equation}
    \Pr(\mbox{difference} \mid \mbox{ungapped, variable}) = \frac{3}{4}\left(1 - e^{-\frac{4t}{3(1-\pin)}}\right)\label{jciProb}.
\end{equation}
The only difference between this formula and the Jukes-Cantor transition probability is the inclusion
    of a $1-\pin$ factor to increase the rate of substitution for the variable sites.
This is included to adhere to the common convention that the mean rate of substitutions is equal to 1.0 per site.

For a pair of sequences, the probability of seeing a different state at a randomly chosen, ungapped, variable
    site (Eqn. \ref{jciProb}) is a monotonically increasing function of $t$.
However, the proportion of ungapped sites which are variable decreases, as was shown in Eqn.~(\ref{propUngappedVar}).
The expected pairwise distance for the paired invariants model
    when measured as the expected proportion of ungapped positions that differ between the tips is:
\begin{eqnarray}
    \mathbb{E}[p] & = & \Pr(\mbox{difference} \mid\mbox{ungapped, variable}) \Pr(\mbox{variable}\mid\mbox{ungapped}) \nonumber\\
                                    & = & \frac{ {3 (1 - \pin) e^{-\theta t} (1-  e^{-\frac{4 t}{3(1-\pin)}}))}}{4\left(\pin + (1-\pin)e^{-\theta t}\right)}. \label{expPPaired}
\end{eqnarray}
This expected $p$-distance is shown in figure \ref{figgen}{B}.
Note that it is not a monotonically increasing function.

\subsection{Gaps-as-missing distance correction}
Under a gaps-as-missing analysis, only the ungapped columns are relevant in distance calculations.
Thus, the expected $p$-distance shown in Eqn.~(\ref{expPPaired}) fills the role of $g(t)$ in the discussion of
    our proofs about the consistency of
    distance-based tree estimation.
Note that the substitution model for the paired invariant sites model is simply the Jukes-Cantor substitution model with invariant sites.
If we assume that we know the (correct) proportion of invariant residues in the generating process, 
    then the distance correction for this model is:
\begin{equation}
    f^{-1}(p) = \frac{-3(1-\pin)}{4}\ln\left(1-\frac{4p}{3(1-\pin)}\right). \label{jcicorr}
\end{equation}

\begin{figure}
\caption{The transformed evolutionary distance $\overline{t}$ values as a function of
true evolutionary distance $t$ (Eqn. \ref{expPCorrected}).}
\includegraphics[scale=0.7]{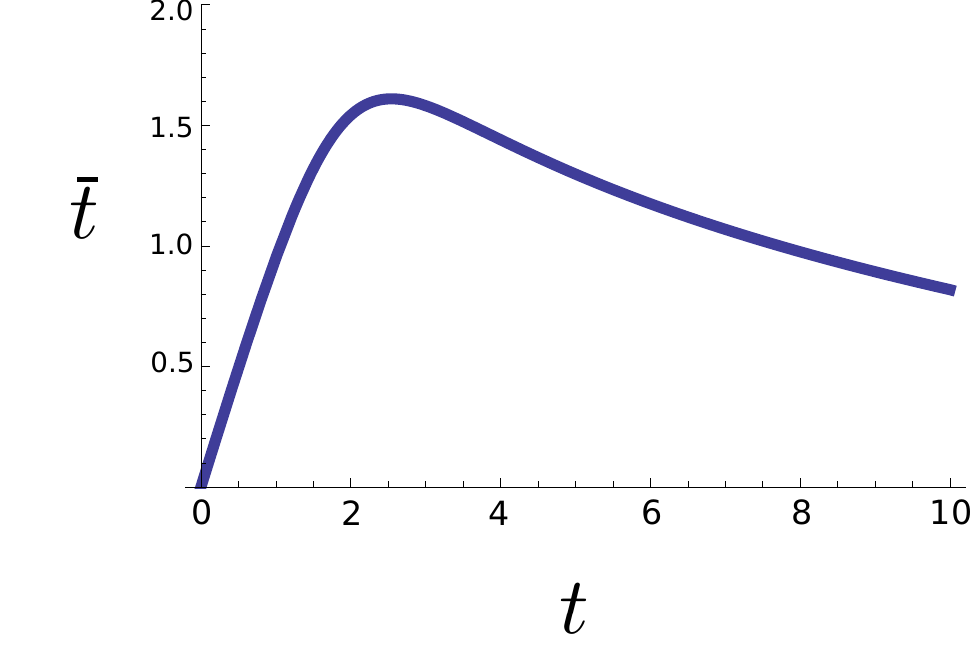}
\label{figcorr}
\end{figure}

We can combine Eqn.~(\ref{expPPaired}) and (\ref{jcicorr}) to express the transformed evolutionary distances
    $\overline{t}$ as a function of the true evolutionary distance, $t$:
\begin{eqnarray}
    \overline{t} & = & \frac{-3(1-\pin)}{4}\ln\left(1-\frac{4\frac{ {(1-\pin)e^{-\theta t} (\frac{3}{4} - \frac{3}{4} e^{\frac{-4 t}{3(1-\pin)}})}}{\pin + (1-\pin)e^{-\theta t}}}{3(1-\pin)}\right) \nonumber\\
      & = & \frac{-3(1-\pin)}{4}\ln\left(1-\frac{{e^{-\theta t} (1 - e^{\frac{-4 t}{3(1-\pin)}})}}{\pin + (1-\pin)e^{-\theta t}}\right) \label{expPCorrected}
\end{eqnarray}
This  function is shown in Figure \ref{figcorr}.

Clearly the function is not linear; indeed it is not monotonically increasing.
In fact, the function is not linear even at small path lengths.
The first and second derivatives of the distance correction with respect to $t$ (see Appendix B)
    are somewhat complicated.
However, when evaluated at $t=0$, the first derivative is 1 and the
second derivative of the expected value of the distance correction is $-2\pin\theta$.
Thus, the gaps-as-missing-data approach coupled with the correct substitution
    model results in a concave distance correction function whenever both $\pin>0$ and $\theta>0$.
Lemma \ref{concaveThm} of Appendix A states that this will lead
    to statistically inconsistent estimation of the tree topology
    for some tree shapes.

\section*{Conclusions}
We have extended the work of \cite{susko_inconsistency_2004} by proving that
    there is a tree shape which will lead to the positively misleading
    estimation of an incorrect tree topology when the distance correction
    function is convex.
We have also proven that the commonly applied gaps-as-missing-data approach
    will not lead to statistical inconsistency of distance estimates if
    the indel and substitution processes are independent.
However, sequence evolution follows the paired invariants model, the deviation
    from independence is sufficient to lead to inconsistency of the distance
    estimates and the tree topology.

Obviously, the paired invariants model with a Jukes-Cantor substitution process
    is an extremely simple model which does not accurately depict the evolution of real
    sequences.
Nevertheless, the paired invariants model encapsulates a simple idea that has been
    at the core of thinking about molecular evolution ever since \citet{Kimura1968}:
    constant sites probably are constrained because they play an important functional
    role.
It seems entirely plausible that the subset of functionally important sites in 
    the genome would be prevented from experiencing fixation of indels or substitutions.
Thus it is troubling that adding this idea to the simplest possible substitution
    model is sufficient to lead to inconsistency of phylogenetic inference.

One obvious solution would be to rely on distance corrections which do not treat gaps 
    as missing data.
Another option may be using multiple values of $\pin$ to correct for the fact that the
    proportion of ungapped positions which correspond to constrained sites is likely
    to be higher for comparisons over long evolutionary timespans.
Both the proportion of gapped sites in the correct pairwise sequence alignment
    and the proportion of ungapped positions which are variable (shown in Figure \ref{figgen}A)
    are monotonically changing functions of the path length.
This implies that it may be possible to devise some recipe for correcting distances
    that uses a pair-specific value of $\pin$, and that this pair-specific
    $\pin$ could be calculated from an observable property of an alignment.
Such a procedure might rescue distance-based from inconsistency when the data are generated
    by the paired invariants model.
However, this form of inference would probably be sensitive to slight inadequacies of the model
    because accounting for rate heterogeneity when using pairwise data alone is notoriously 
    difficult.
Our results underscore that fact that phylogenetic inference is a problem that is so difficult
    that even subtle forms of ascertainment bias can lead to fundamental misbehavior of
    inference methods.
    
\section*{Acknowledgements}
MTH was supported by ATOL-0732920, the AVAToL Open Tree of Life award and HITS. EJM was supported by an Alexander von Humboldt award.

\bibliography{distanceconsistency}

\begin{thebibliography}{}

\bibitem[Felsenstein, 1978]{felsenstein_cases_1978}
Felsenstein, J. (1978).
\newblock Cases in which parsimony or compatibility methods will be positively
  misleading.
\newblock {\em Systematic Biology}, 27(4):401--410.

\bibitem[Grievink et~al., 2013]{grievink_missing_2013}
Grievink, L.~S., Penny, D., and Holland, B.~R. (2013).
\newblock Missing data and influential sites: Choice of sites for phylogenetic
  analysis can be as important as taxon sampling and model choice.
\newblock {\em Genome Biology and Evolution}, 5(4):681--687.

\bibitem[Jukes and Cantor, 1969]{jukes_evolution_1969}
Jukes, T.~H. and Cantor, C.~R. (1969).
\newblock Evolution of protein molecules.
\newblock pages 21--132. Elsevier.

\bibitem[Kimura, 1968]{Kimura1968}
Kimura, M. (1968).
\newblock Evolutionary rate at the molecular level.
\newblock {\em Nature}, 217(5129):624--626.

\bibitem[Roure et~al., 2013]{roure_impact_2013}
Roure, B., Baurain, D., and Philippe, H. (2013).
\newblock Impact of missing data on phylogenies inferred from empirical
  phylogenomic data sets.
\newblock {\em Molecular Biology and Evolution}, 30(1):197--214.

\bibitem[Saitou and Nei, 1987]{saitou_neighbor-joining_1987}
Saitou, N. and Nei, M. (1987).
\newblock The neighbor-joining method: a new method for reconstructing
  phylogenetic trees.
\newblock {\em Molecular Biology and Evolution}, 4(4):406--425.

\bibitem[Soubrier et~al., 2012]{soubrier2012}
Soubrier, J., Steel, M., Lee, M., Der~Sarkissian, C., Guindon, S., Ho, S.,
  Cooper, A., and the Genographic~Consortium (2012).
\newblock The influence of rate heterogeneity among sites on the time
  dependence of molecular rates.
\newblock {\em Molecular {B}iology and {E}volution}, 29(5):3345--3358.

\bibitem[Susko et~al., 2004]{susko_inconsistency_2004}
Susko, E., Inagaki, Y., and Roger, A.~J. (2004).
\newblock On inconsistency of the neighbor-joining, least squares, and minimum
  evolution estimation when substitution processes are incorrectly modeled.
\newblock {\em Molecular Biology and Evolution}, 21(9):1629--1642.

\bibitem[Tajima, 1993]{tajima93}
Tajima, F. (1993).
\newblock Unbiased estimation of evolutionary distance between nucleotide
  sequences.
\newblock {\em Molecular Biology and Evolution}, 10(3):677--688.

\bibitem[Thorne et~al., 1991]{thorne_evolutionary_1991}
Thorne, J.~L., Kishino, H., and Felsenstein, J. (1991).
\newblock An evolutionary model for maximum likelihood alignment of {DNA}
  sequences.
\newblock {\em Journal of Molecular Evolution}, 33(2):114--124.

\bibitem[Warnow, 2012]{Warnow2012}
Warnow, T. (2012).
\newblock Standard maximum likelihood analyses of alignments with gaps can be
  statistically inconsistent.
\newblock {\em PLoS Currents}, 4.

\end{thebibliography}

\section*{Appendix A}
Suppose that distances are generated on a tree by a model $M_g$ and corrected
assuming a model $M_f$.

\begin{theorem}
\label{mainThm}
\mbox{ }
Suppose that $f(t)$ and $g(t)$ (the functions for correcting and generating $p$-distances respectively) are analytic functions of $t$ that are strictly increasing  in some neighbourhood of 0, and satisfy $f(0)=g(0) =0$. Let $h(t) = \overline{t} = f^{-1}(g(t))$ (the transformed evolutionary distances).  Then precisely one of the following conditions holds:
\begin{itemize}
\item
The correction process $f$ is equal to the generating function $g$ up to a scalar multiple (i.e.
$f(t) = g(t/c)$ and so $h(t)=ct$ for all $t \in [0,\rho)$, for some constant $c$). In this  case NJ will select the correct tree topology when applied to the transformed evolutionary distances; or 
\item
The correction process $f$ is not equal to the generating function $g$ up to a scalar multiple.  In this case there exists a binary tree on four leaves with an associated set of strictly positive branch lengths for which NJ will select an incorrect tree topology when  applied to the transformed distances. 
\end{itemize}
\end{theorem}

The proof of this result involves combining five lemmas;
    the first is standard,
    the second is a formal statement of results from \citet{susko_inconsistency_2004},
    the third is new,
    and the fourth and fifth are technical lemmas.
\begin{lemma}
\label{sai}
[\citet{saitou_neighbor-joining_1987} p.413]
NJ applied to distance data on four taxa ($A,B,C,D$) returns the quartet tree $AB|CD$ if 

$d_{AB}+d_{CD} < \min\{d_{AC}+d_{BD}, d_{AD}+d_{BC}\}$.
\end{lemma}

\begin{lemma}
\label{concaveThm}
Suppose the transformed distance  function $h(t)$ is strictly concave and increasing on the interval $[\lambda, 2\lambda]$ for some $\lambda>0$.
For any $\sigma>0$ sufficiently small, distances on Felsenstein-Zone tree of Fig.~\ref{figcom}(A) that are transformed by $h$ have the property that NJ will 
estimate the incorrect tree topology ($AD|BC$).
\end{lemma}

{\em Proof of Lemma~\ref{concaveThm}.}

By Lemma~\ref{sai}, for  any distance function $d$  on four taxa $i,j,k,l$, NJ applied to $d$ will  return the quartet tree $ij|kl$
when $i,j$ minimizes the pairwise sum $d_{ij}+d_{kl}$. Let us now put $d_{ij} =h(t_{ij}) =  \overline{t}_{ij}$ (i.e. the transformed evolutionary distances). Consider the three pairwise sums:
\begin{itemize}
\item[(S1)]  $d_{AB}+d_{CD} = 2h(\lambda+ \sigma);$
\item[(S2)]  $d_{AC}+d_{BD} = 2h(\lambda+2\sigma);$
\item[(S3)]  $d_{AD}+d_{BC} = h( 3\sigma)+ h(2\lambda+\sigma) ;$
\end{itemize}
Since $h$ is strictly increasing on $[\lambda, 2\lambda]$, the expression (S2) is always greater than (S1) for any $\sigma>0$.
 Thus it suffices to show that case (S3), which corresponds to NJ returning the tree $AD|BC$, is less that (S1) for sufficiently small $\sigma>0$.
To this end, note that since $h$ is strictly concave on $[\lambda, 2\lambda]$ we have:
$h(2\lambda) < 2h(\lambda),$
so if we let $$q(x) :=2 h(\lambda +x) -h(2\lambda+x)-h(3x)$$
then $q(0) = 2h(\lambda) - h(2\lambda) -h(0) >0$
(recall $h(0)=0$).  Since $h$ is continuous (by virtue of being analytic) $q$ is too, so it follows that for any sufficiently small (but strictly positive) value of $\sigma$ we have 
$q(\sigma)>0$.
Because $q(\sigma)$ equals the quantity described by (S1) minus that described by (S3), when $q(\sigma) > 0$ then
NJ will prefer tree $AD|BC$ over the true tree $AB|CD$.
\hfill$\Box$

\begin{lemma}
\label{convexThm}
Suppose the transformed distance function $\overline{t}=h(t)$ is strictly convex and increasing on the interval $[\lambda, 2\lambda]$ for some $\lambda>0$.
For any $\sigma>0$ sufficiently small, distances on the `twisted' Farris-Zone tree of Fig.~\ref{figcom}(B) that are transformed by $h$ have the property that neighbor-joining will 
estimate the incorrect tree topology ($AD|BC$).
\end{lemma}

{\em Proof of Lemma~\ref{convexThm}.}
For the `twisted' Farris-Zone tree of Fig.~\ref{figcom}(B)  consider the three pairwise sums:
\begin{itemize}
\item[(S1)]  $d_{AB}+d_{CD} = h(2\lambda+ \sigma)+ h(3\sigma);$
\item[(S2)]  $d_{AC}+d_{BD} = h(\lambda+4\sigma)+ h(\lambda + 2\sigma);$
\item[(S3)]  $d_{AD}+d_{BC} =2 h( \lambda + 3\sigma).$
\end{itemize}

Now, if $h$ is strictly convex on $[\lambda, 2\lambda]$ and if $x,y \in [\lambda, 2\lambda]$ then:
$$h\left( \frac{x+y}{2} \right) < \frac{1}{2}[h(x)+h(y)].$$
Applying this with $x = \lambda + 4\sigma$ and $y= \lambda + 2\sigma$, where $0< \sigma < \lambda/4$, gives:
$$h(\lambda + 3\sigma) < \frac{1}{2}[h(\lambda + 4\sigma) + h(\lambda + 2\sigma)],$$
which gives (S3)$<$(S2). 

Again by convexity, $h(2\lambda)> 2h(\lambda)$, 
so if we let $$q(x) := h(2\lambda+ \sigma)+ h(3\sigma) - 2 h( \lambda + 3\sigma).$$
then $q(0) = h(2\lambda) - 2h(\lambda) +h(0) >0$.
By a similar continuity argument as in the concave case, there exists $\sigma>0$ so that $q(\sigma)>0$.
Because $q(\sigma)$ equals the quantity described  (S1) minus that described by (S3), conditions for which $q(\sigma) > 0$ are conditions
for which NJ will again prefer tree $AD|BC$ over the true tree $AB|CD$.

\hfill$\Box$

\begin{lemma}
\label{analyticlem}
Under the assumptions on $f$ and $g$ in Theorem~\ref{mainThm}, the transformed distance function $h$ is a strictly increasing analytic function of $t$ on $[0, \rho)$ for some $\rho>0$.
\end{lemma}
{\em Proof of Lemma~\ref{analyticlem}.}
The proof that $h$ is analytic is straightforward, since analytic functions (in particular $f$ and $g$) are closed under composition, and also under functional inverse (providing their derivative  is non-zero, as it is here). To see that $h$ is increasing, at least close to 0,  note that, by elementary differential calculus, we have:
\begin{equation}
\label{ddt}
\frac{d}{dt}h(t) = \frac{g'(t)}{f'(f^{-1}(g(t)))}.
\end{equation}
By assumption, $f$ and $g$ are both increasing in some neighbourhood of $0$, and since $f^{-1}(g(0)) =f^{-1}(0)=0$,  there exists $\rho>0$ for which the numerator and denominator of Inequality (\ref{ddt}) are both strictly positive for all  $t \in [0, \rho)$. 
\hfill$\Box$

\begin{lemma}
\label{taylorThm}
Suppose $H(t)$ is a real-valued function that is analytic in $[0, \rho)$ for some $\rho>0$, and that satisfies $H(0)=0$.
  If $H(t) \neq ct$ on $[0, \rho)$ for some constant $c$, then there exists some value $s>0$
so that $H(t)$ is either strictly concave on the interval $[s/2, s]$ or strictly convex on the  interval $[s/2, s]$. 
\end{lemma}

{\em Proof of Lemma~\ref{taylorThm}.}

If $H''(0)>0$ then since $H''$ is continuous at $0$, there is a value $s \in [0, \rho)$ so that $H''(t)>0$ for all $t\in [0, s]$ and so $H$ is strictly convex on $[s/2,s]$.
Similarly, if $H''(0)<0$ then $H$ is strictly concave on $[s/2, s]$ for some $s>0$.
Suppose that $H''(t)=0$.  Then either (i)  there exists a smallest $k>2$ for which $H^{(k)}(0) \neq 0$ (call this value $k_1$) or (ii) $H^k(0)=0$ for all $k>2$. In Case (i), suppose first that $a:=H^{(k_1)}(0) >0$. 
A Taylor series expansion of $H$ about 0 gives $H(t)= at^{k_1}+\cdots$ where the remaining terms are of order $t^{k+1}$ and higher.
Thus, for a sufficiently small $\nu \in (0, \rho)$,  we have $H''(t)= k_1(k_1-1)at^{k_1-2} +  (\mbox{terms of order $t^{k_1-1}$ and higher})$ so $H''(t) >0$ for all $t \in (0, \nu)$. In particular,  for any strictly positive value of $s$ less than $\nu$ we have $H''(t)>0 $ for all $t \in [s/2, s]$. 
Thus, as before, $H''$ is strictly convex on $[s/2,s]$. 
A similar argument (for strict concavity) applies if $H^{(k_1)}(0) <0$.
In Case (ii) the Taylor expansion of $H(t)$ on $[0, \rho)$ centered on 0, shows that $H''(t)=0$ for all $t \in [0, \rho).$  
By integrating (twice) it follows that $H(t) = ct + H(0)$ for all $t \in [0, \rho)$, for some constant $c$.  Since $H(0)=0$, this gives $H(t) = ct$, as claimed.  
\hfill$\Box$

\bigskip

{\em Proof of  Theorem~\ref{mainThm}:}  By Lemma~\ref{analyticlem}, $h$ and analytic and  increasing in $[0, \rho)$, so by Lemma~\ref{taylorThm}, if we take $H(t) = h(t)$ then if $h$ is not linear, it is either strictly concave or strictly convex on an interval of the form $[s/2, s]$ for some $s \in (0, \rho)$.  Theorem~\ref{mainThm} now follows from Lemmas~\ref{concaveThm} and \ref{convexThm}.
\hfill$\Box$

\section*{Appendix B}
The first and second derivatives of the expected corrected distance
(equation \ref{expPCorrected}) with respect to the path length $t$ are:
\begin{eqnarray}
    c[t] & = & e^{\frac{4 t}{3 - 3\pin}} \nonumber\\
    d[t] & = & e^{t \left(\theta + \frac{4}{3 - 3\pin}\right)} \nonumber\\
    \frac{\partial \overline{t}}{\partial t} & = & \frac{
            4
            - 3(e^{\theta t} - d[t])\pin^{2}\theta 
            + \pin\left(
                    e^{\theta t}\left[4 + 3 \theta\right]
                    -4
            -3 d[t]\theta \right)}{
          4
          \left(1 - c[t]\pin + d[t]\pin\right)
          \left(1 + e^{\theta t} \pin - \pin\right)
      }  \nonumber\\
    v[t] & = & e^{t \left(\theta + \frac{8}{3 - 3\pin}\right)} \nonumber\\
    w[t] & = & e^{t \left(3\theta + \frac{8}{3 - 3\pin}\right)} \nonumber\\
    x[t] & = & e^{t \left(3\theta + \frac{4}{3 - 3\pin}\right)} \nonumber\\
    y[t] & = & e^{2t \left(\theta + \frac{2}{3 - 3\pin}\right)} \nonumber\\
      m & = & \pin - 1 \nonumber \\
      u[t] & = &  -16c[t] m^2
        + 9e^{\theta t} m^3 \theta^2
        + 9 v[t] m^3 \pin \theta^2
        -9 w[t] m^2 \pin^2 \theta^2 \nonumber \\
    & & + x[t] \pin^2\left(4 - 3m\theta\right)^2
        + 16 y[t]\pin\left(2 + 3 \theta + 3 \pin^2 \theta - 3 \pin[1 + 2\theta]\right)\nonumber\\
        & & -d[t]m\left(3\pin^2(8-3\theta)\theta 
                + 9\pin^3\theta^2
                + (4+3\theta)^2
                - 3 \pin[16 + 16\theta + 3 \theta^2]
                \right) \nonumber \\
    \frac{\partial^2 \overline{t}}{\partial t^2} & = & \frac{
      \pin u[t]
  }{
      12m
      \left(1 - c[t]\pin + d[t]\pin\right)^2
      \left(1 + e^{\theta t} \pin - \pin\right)^2
  } \nonumber
\end{eqnarray}
These were calculated using the Mathematica notebook included as part of the supplementary materials.

\end{document}